\documentclass[twocolumn,aps,prl,showpacs,floatfix,showkeys]{revtex4}

\usepackage{amsfonts}
\usepackage[english]{babel}
\usepackage{amsmath,amssymb}
\usepackage[all,cmtip]{xy}

\def\be{\begin{equation}}
\def\ee{\end{equation}}
\def\ba{\begin{eqnarray}}
\def\ea{\end{eqnarray}}
\begin{document}

\title{Reduction of superintegrable systems: the anisotropic harmonic
oscillator}
\author{Miguel A. Rodr\'{\i}guez}\email[ E-mail: ]{rodrigue@fis.ucm.es}
\affiliation{Departamento de F\'{\i}sica Te\'{o}rica II, Facultad de F\'{\i}sicas,
Universidad Complutense, 28040 -- Madrid, Spain}
\author{Piergiulio Tempesta}\email[E-mail: ]{p.tempesta@fis.ucm.es}
\affiliation{Departamento de F\'{\i}sica Te\'{o}rica II, Facultad de F\'{\i}sicas, Universidad
Complutense, 28040 -- Madrid, Spain}
\author{Pavel Winternitz}\email[ E-mail: ]{wintern@crm.umontreal.ca}
\affiliation{Centre de recherches math\'{e}matiques, Universit\'{e} de Montr\'{e}al, C.P.
6128, Succursale centre--ville, Montr\'{e}al (QC) Canada}
\keywords{superintegrability, classical anisotropic oscillators}
\pacs{45.20.Jj, 02.30.Ik}

\begin{abstract}
We introduce a new $2N$--parametric family of maximally superintegrable systems in $N$ dimensions, obtained as a reduction of an anisotropic harmonic oscillator in a $2N$--dimensional configuration space. These systems possess closed bounded orbits and integrals of motion which are polynomial in the momenta. They generalize known examples of superintegrable models in the Euclidean plane.
\end{abstract}

\volumeyear{year}
\volumenumber{number}
\issuenumber{number}
\eid{identifier}
\date{June 16, 2008}
\startpage{1}
\endpage{6}
\maketitle

\section{I. Introduction}

The aim of this paper is to introduce a new class of maximally superintegrable systems that are obtained as a symplectic reduction of the anisotropic harmonic oscillator. These systems depend on a set of $N$ real and $N$ integer parameters and possess integrals of motion polynomial in the momenta.
The Hamiltonian defining this family is
\begin{equation}
H_{N}=\frac{1}{2}\sum_{i=1}^{N}p_{i}^{2}+\frac{1}{2}\sum_{i=1}^{N}\frac{k_{i}}{x_{i}^{2}%
}+\frac{\omega ^{2}}{2}\sum_{i=1}^{N}n_{i}^{2}x_{i}^{2}  \label{I.a}
\end{equation}%
We recall that in classical mechanics, superintegrable (also known as noncommutatively integrable \cite{MF}) systems are characterized by the fact that they possess more than $N$ integrals of
motion functionally independent, globally defined in a $2N$--dimensional phase space. In particular, when the number of integrals is $2N-1$, the systems are said to be maximally superintegrable.
The dynamics of these systems is particularly interesting: all bounded orbits are closed and periodic. This issue, for the spherically symmetric potentials, was first noticed by Bertrand \cite%
{Bertrand}. The phase space topology is also very rich: it has the structure
of a symplectic bifoliation, consisting of the usual Liouville--Arnold
invariant fibration by Lagrangian tori and of a (coisotropic) polar
foliation \cite{Nekh}, \cite{Fasso}. Apart from the harmonic oscillator and
the Kepler potential, many other potentials turn out to be superintegrable,
like the Calogero--Moser potential, the Smorodinsky--Winternitz systems, the
Euler top, etc.

A considerable effort has recently been devoted to the search
for superintegrable systems as well as to the study of the algebraic and
analytic properties of these models. For a recent review of the topic, see
\cite{TWR}.

The notion of superintegrability possesses an interesting analog in quantum
mechanics. Sommerfeld and Bohr were the first to notice that systems
allowing separation of variables in more than one coordinate system may
admit additional integrals of motion. Superintegrable systems show
accidental degeneracy of the energy levels, which can be removed by taking
into account the quantum numbers associated to the additional integrals of
motion. One of the best examples of this phenomenon is provided by the
Coulomb atom \cite{Fock}, \cite{Bargmann}, \cite{JH}, which is
superintegrable in $N$ dimensions \cite{Englefield}, \cite{RW}. A systematic search for quantum
mechanical potentials exhibiting the property of superintegrability was
started in \cite{FMSUW}, \cite{MSVW} and \cite{Winternitz}. These models in
many cases are also exactly solvable, i.e. they possess a spectrum
generating algebra, which allows to compute the whole energy spectrum
essentially by algebraic manipulations \cite{TTW}. In classical
mechanics, the multiseparability of the Hamilton--Jacobi equation implies
that there should exist at least two different sets of $N$ quadratic integrals of motion in involution.
Reduction techniques both in classical and quantum mechanics are well--known (see, for instance, \cite{CCM}).
Essentially, the common idea of several of the existing approaches is to start from a
free motion Hamiltonian defined in a suitable higher--dimensional space and
to project it down into an appropriate subspace. In this way, one gets a
reduced Hamiltonian that is no longer free: an integrable potential appears
in the lower--dimensional space \cite{ORW}. A different point of view, that we adopt here, is to
start instead directly from a nontrivial (i.e. not free) dynamical system in
a given phase space and to reduce it to a proper subspace, in such a way
that the superintegrability of the considered system be inherited by the reduced one.

In this work, we study the reduction of an anisotropic
harmonic oscillator, defined in a $2N$--dimensional classical configuration space. This system is
maximally superintegrable. It is described by the Hamiltonian%
\begin{equation}
H_{2N}=\frac{1}{2}\sum_{i=1}^{2N}\hat{p}_{i}^{2}+\frac{\omega
^{2}}{2}\sum_{i=1}^{2N}n^2_{i}y_{i}^{2}\text{.}  \label{I.1}
\end{equation}%
We will prove that it can be suitably reduced to the system (\ref{I.a}), and that this new system is still maximally superintegrable, with integrals
of motion inherited from the system (\ref{I.1}). This goal is
achieved under the assumption $n_{1}=n_{2},...,n_{2N-1}=n_{2N}$.
From a geometrical point of view, the approach we adopt reposes on the Marsden--Weinstein symplectic reduction scheme \cite{AM}, \cite{LM}, \cite{MW}.
Given a symplectic manifold $(M, \Omega)$, let $K_{1},...,K_{k}$ be $k$ functions in involution:
\begin{equation}
\{K_{i}, K_{j} \} = 0 \hspace{5mm} i,j = 1,...,k.
\end{equation}%
Assume also that $\mathrm{d}K_{i}$ are independent at each point. Since the flows of the associated Hamiltonian vector fields $X_{K_{1}},...,X_{K_{k}}$ commute, they can be used to define a symplectic action of $G=\mathbb{R}^{k}$ on the manifold. Let $J$ be the momentum map of this action, and $\mu$ be a regular value for $J$. Then we can conclude that $P_{\mu}:= J^{-1}(\mu)/G$ is still a symplectic manifold, of dimension $\dim M - 2k$, called the reduced phase space.
In our case, $K_{i}, i=1,...,N$ are components of the angular momentum, $J = K_{1}\times...\times K_{N}$ is the momentum map, $G=SO(2)\times SO(2) \times...\times SO(2)$ ($N$ times), and the reduced space is $P_{\mu}=J^{-1}$ $(\mu)/T^{N}$, where $T^{N}$ is the $N$--dimensional torus and $\dim P_{\mu} = 2N$. This procedure is a generalization of what in celestial mechanics, since the work of Jacobi, is called "elimination of the nodes" (see \cite{AM}, chapter IX for details).
The reduced Hamiltonian is reminiscent of the structure of the original Hamiltonian, defined in the
$4N$--dimensional phase space, but also possesses a Rosochatius--type term \cite{Rosochatius}, \cite{GHWH}, involving
parameters $k_{i}$ corresponding to the variables that become ignorable, in addition to the harmonic part. Therefore, using the reduction procedure, we obtain the parametric family of Hamiltonian systems (\ref{I.a}), defined on a reduced phase $P_{\mu}$.

The transformations we consider, although very simple, are non--trivial, since the reduced Hamiltonian is not shape--invariant. Nevertheless, since the reduced system turns out to be maximally superintegrable, bounded orbits still remain closed in the reduced space.

For $N=2$ maximal superintegrability  \cite{FMSUW}, \cite{Winternitz} and exact solvability \cite{TTW} of the system (\ref{I.a}) was already established for $n_1=n_2=1$ and for $n_1=1,n_2=2,k_2=0$. The integrals of motion in these cases are second order in the momenta.

Here we will show that in the general case ($n_i$ and $N$ arbitrary positive integers and $k_i$ arbitrary real numbers) $N$ integrals can be chosen to be of order two, the other $N-1$ functionally independent ones of order $n_i+n_N$ or $n_i+n_N-1$. Other systems possessing third and higher order integrals have been studied in the literature   \cite{Drach}, \cite{FL}, \cite{Gravel}, \cite{GW}, \cite{Ra}, \cite{Tsiganov}.

This paper is directly related to the recent interesting work by Verrier and
Evans \cite{Verrier}, that performed a similar reducing transformation for
the Kepler potential. They found a superintegrable system in three
dimensions possessing a quartic integral. They also conjectured that the
system (\ref{I.a}) in three space dimensions should be maximally
superintegrable, although the explicit expression of the integrals remained
to be determined. In the following, we will prove this conjecture, and also
we will establish that the system (\ref{I.a}) is maximally superintegrable
in full generality, i.e. for $N$ arbitrary, providing explicitly the
corresponding set of integrals of the motion.

After the present article was submitted we learned of a new article by Evans and Verrier \cite{EvansII} in which the authors also establish the superintegrability of the system (\ref{I.a}) for $N=3$. Their results are compatible with ours, though they express the integrals in terms of Chebyshev polynomials. Moreover, they also treat the quantum analogue of system (\ref{I.a}) and establish the degeneracy of the energy levels related to the representation theory of the group $SU(3)$.

The paper is organized as follows. In Section II, the main properties of the
anisotropic oscillator are briefly reviewed. Then its reduction to the
planar case is studied in detail. We will show how superintegrability is
preserved under a multipolar change of variables and subsequent reduction. In Section III, the same
problem is treated and solved in full generality. Some open problems are
discussed in the final Section.

\section{II. Reduction of the anisotropic oscillator}

The anisotropic oscillator in the two--dimensional case both in classical
and quantum mechanics was discussed by Jauch and Hill \cite{JH}, \cite{Demkov}, \cite{Ilkaeva}. The system (%
\ref{I.1}) is also known to be superintegrable in any dimension, if the ratios
of the frequencies are rational. Let us consider a $2N$ dimensional space and assume
\begin{equation}
\frac{\omega _{1}}{n_{1}}=\frac{\omega _{2}}{n_{2}}=...=\frac{\omega _{2N}}{n_{2N}}=\omega ,\qquad n_{i}\in \mathbb{N} \label{II.1}
\end{equation}%
Following \cite{JH}, we define the set of invariants in an auxiliary complex phase
space, with coordinates $z_{i},\bar{z}_{i}$, $i=1,...,2N$. Precisely,%
\begin{equation}
z_{j}=\hat{p}_{j}-i n_{j}\omega y_{j},\qquad\overline{z}_{j}=\hat{p}_{j}+i n_{j}\omega
y_{j}.  \label{II.2}
\end{equation}%
It is easily checked that the expressions%
\begin{equation}
c_{jk}=z_{j}^{n_{k}}\bar{z}_{k}^{n_{j}}  \label{II.3}
\end{equation}%
provide integrals of motion. They can be also arranged in a real--valued
form, as the combinations $(1/2)\left( c_{ij}+\overline{c}_{ij}\right) $ and $%
(1/2i)\left( c_{ij}-\overline{c}_{ij}\right) $. In particular, among these
integrals we have the angular momenta
\begin{equation}
L_{ik}=y_{i}\hat{p}_{k}-y_{k}\hat{p}_{i}  \label{II.4}
\end{equation}%
(when $n_{i}=n_{k}$) and the tensor
\begin{equation}
T_{ik}=\hat{p}_{i}\hat{p}_{k}+n_{i} n_{k}  \omega^2  y_{i}y_{k}  \label{II.5}
\end{equation}%
We will now study reductions of the anisotropic oscillator (\ref{I.1}) and establish the
superintegrability of the corresponding dynamical systems.

\subsection{Hamiltonian and first integrals: the planar case}

We recall the definition of a momentum map. For further details, see for instance \cite{AM}. Let $(M, \Omega)$ be a $2n$ dimensional symplectic manifold. Suppose that a Lie group $G$ acts on $M$ and leaves $\Omega$ invariant. Let $\mathfrak{g}$ be the Lie algebra of $G$, $\mathfrak{g^*}$ its dual space, and $<,>$ the natural pairing between the two spaces.

A momentum map for the $G$--action on $(M,\Omega)$ is a map $J: M\rightarrow \mathfrak{g^{*}}$ such that, for all $X\in\mathfrak{g}$,
$$
d(\langle J,X\rangle)=i_{X}\Omega
$$
In particular, if the manifold is exact, i.e. $\Omega = d \theta$, and the $G$-action leaves $\theta$ invariant as well, we have
$$
J_{X}=i_{X}\theta
$$
We will also assume that the map is equivariant with respect to the coadjoint action $\mathrm{Ad}^{*}$ of $G$ on $\mathfrak{g^{*}}$, i.e.:
$$
\langle  \mathrm{Ad}^{*}_g \,\xi,X \rangle  = \langle \xi, \mathrm{Ad}_{g^{-1}}X \rangle,
$$
for all $g\in G, \xi\in\mathfrak{g^{*}}$ and $X\in\mathfrak{g}$.

Let us first consider a simple case, when the anisotropic oscillator is defined in a symplectic manifold $M$ with $\dim M=4$. So, $\Omega=\sum_{i=1}^{4}dy^{i}\wedge d\hat{p}_{i}$.
In order to make the reduction possible, we select frequencies to be equal in pairs, so that we have only two independent frequencies. Hence the system (\ref{I.1}) takes the special form

\begin{equation}
H_{4}=\frac{1}{2}(\hat{p}_{1}^{2}+\hat{p}_{2}^{2}+\hat{p}_{3}^{2}+\hat{p}%
_{4}^{2})+\frac{n_{1}^{2}\omega ^{2}}{2}(y_{1}^{2}+y_{2}^{2})+\frac{%
n_{2}^{2}\omega ^{2}}{2}(y_{3}^{2}+y_{4}^{2})\text{.}  \label{3.1}
\end{equation}%
In the auxiliary coordinates $z_{1}$,$\bar{z}_{1}$,...,$z_{4}$,$\bar{z}_{4}$%
, we have explicitly
\begin{eqnarray}
\nonumber z_{1} &=&\hat{p}_{1}-{\mathrm{i}}\,n_{1}\omega y_{1},\quad z_{2}=\hat{p}_{2}-%
{\mathrm{i}}\,n_{1}\omega y_{2},\quad  \label{3.2} \\
z_{3} &=&\hat{p}_{3}-{\mathrm{i}}\,n_{2}\omega y_{3},\quad z_{4}=\hat{p}_{4}-%
{\mathrm{i}}\,n_{2}\omega y_{4}\text{.}
\end{eqnarray}%
Consequently, the Hamiltonian reads
\begin{equation}\label{Ham}
H_{4}=\frac{1}{2}\sum_{i=1}^{4}\text{ }|z_{i}|^{2}\text{.}
\end{equation}%
Put in a matrix form, the set of invariants (\ref{II.3}) can be represented by the matrix
\begin{equation}\label{matice}
Z=\left(
\begin{array}{cccc}
z_{1}\bar{z}_{1} & z_{1}\bar{z}_{2} & z_{1}^{n_{2}}\bar{z}_{3}^{n_{1}} &
z_{1}^{n_{2}}\bar{z}_{4}^{n_{1}} \\
z_{2}\bar{z}_{1} & z_{2}\bar{z}_{2} & z_{2}^{n_{2}}\bar{z}_{3}^{n_{1}} &
z_{2}^{n_{2}}\bar{z}_{4}^{n_{1}} \\
z_{3}^{n_{1}}\bar{z}_{1}^{n_{2}} & z_{3}^{n_{1}}\bar{z}_{2}^{n_{2}} & z_{3}%
\bar{z}_{3} & z_{3}\bar{z}_{4} \\
z_{4}^{n_{1}}\bar{z}_{1}^{n_{2}} & z_{4}^{n_{1}}\bar{z}_{2}^{n_{2}} & z_{4}%
\bar{z}_{3} & z_{4}\bar{z}_{4}%
\end{array}%
\right) \text{.}
\end{equation}%

Let us consider now the following change of coordinates:
\begin{equation}
\begin{cases} y_{1}=x_{1}\cos x _{3},\quad y_{2}=x_{1}\sin x _{3}\\
y_{3}=x_{2}\cos x _{4},\quad y_{4}=x_{2}\sin x _{4}.
\label{polar}
\end{cases}\end{equation}%
The corresponding momenta read
\begin{align}
& \hat{p}_{1}=-p_{3}\frac{\sin x_{3}}{x_{1}}+p_{1}\cos x_{3},\quad \hat{p}%
_{2}=p_{3}\frac{\cos x_{3}}{x_{1}}+p_{1}\sin x_{3}, \notag  \\
& \hat{p}_{3}=-p_{4}\frac{\sin x_{4}}{x_{2}}+p_{2}\cos x_{4},\quad \hat{p}%
_{4}=p_{4}\frac{\cos x_{4}}{x_{2}}+p_{2}\sin x_{4}.\label{polarmomenta}
\end{align}%

The group  $T_2$, which is the group $SO(2)\times SO(2)$ in the old coordinates, acts on $\mathbb{R}^4$ as follows
\begin{align}
x'_1=&x_1\notag \\
x'_2=&x_2 \notag \\
x'_3=&x_3+a_1 \label{rotace} \\
x'_4=&x_4+a_2. \notag
\end{align}
This group leaves $\Omega$ invariant. The fundamental vector fields on $T^*\mathbb{R}^4$ corresponding to this action are:
\be
X_1=\partial_{x_3},\quad X_2= \partial _{x_4}
\ee
and, if $X=\lambda_1X_1+\lambda_2X_2$, the momentum map $J$ satisfies:
\be
J_{(a_1,a_2)}=\theta(\lambda_1\partial_{x_3}+\lambda_2\partial _{x_4})=\lambda_1p_3+\lambda_2p_4
\ee
Let us choose a regular point in $\mathfrak{t_2^*}$ (the dual of the Lie algebra of $T_2$), for instance
\be\label{17prime}
p_3=\sqrt{k_1},\quad  p_4=\sqrt{k_2}.
\ee
The inverse image under $J$ is
\be
J^{-1}(\sqrt{k_1},\sqrt{k_2})=(p_1,p_2,\sqrt{k_1},\sqrt{k_2},x_1,x_2,x_3,x_4)
\ee

The stabilizer of this point in $t^*_2$ under the coadjoint action of $T_2$ is the whole group, because its action is trivial on the $p$ coordinates.

The reduced phase space is therefore
\be\label{redspace}
J^{-1}(\sqrt{k_1},\sqrt{k_2})/T_2\approx \{(p_1,p_2,x_1,x_2)\in\mathbb{R}^4\}
\ee
and the reduced Hamiltonian is
\be\label{c}
H_2=\frac{p_1^2}{2}+\frac{p_2^2}{2}+  \frac{k_1}{2x_1^2}+ \frac{k_2}{2x_2^2}+\frac{n_1^2}{2}\omega^2 x_1^2+\frac{n_2^2}{2}\omega^2 x_2^2
\ee

Let ${F}$ be a first integral of the Hamiltonian $H_4(\hat{p},y)$, i.e. $\{H_4,F\}=0$.

We show now how the original ring of integrals can be reduced in the low--dimensional phase space.
First, we consider the restriction $\hat{F}$ of the function $F$ to the manifold $J^{-1}(\sqrt{k_1},\sqrt{k_2})$.

Observe that $\hat{F}$ can be defined on the quotient manifold
$J^{-1}(\sqrt{k_1},\sqrt{k_2})/T_2$, when it is constant on the equivalence classes, that is, $\hat{F}$ is independent on $x_3,x_4$. In this case $\hat{F}$ can be factored out in the following way:
$$
\xymatrix{
J^{-1}(\sqrt{k_1},\sqrt{k_2})\ar[r]^{\qquad\qquad\hat{F}}\ar[d]_{\pi} &  \mathbb{R}  \\  J^{-1}(\sqrt{k_1},\sqrt{k_2})/T_2\ar[ur]_{F_r}}
$$
where $\pi$ is the canonical projection and
\be
F_r\circ \pi =\hat{F}.
\ee
Then,
\be
\{H_2,F_r\}=0.
\ee

The integrals of the system (\ref{Ham}) are given in the matrix (\ref{matice}) (though only 7 of them can be functionally independent).
Those that will survive the reduction (\ref{redspace}) are the ones that are left invariant by the $SO(2)\times SO(2)$ rotations (\ref{rotace}). They must Poisson commute with

\begin{align}
L_{12}=& \frac{{\mathrm{i}}}{2n_1\omega}(z_{1}
\bar{z}_{2}-z_{2}\bar{z}_{1})=y_{1}\hat{p}_{2}-y_{2}\hat{p}_{1}, \notag \\
L_{34}=& \frac{{\mathrm{i}}}{2n_2\omega}(z_{3}\bar{z}_{4}-z_{4}\bar{z}_{3})=y_{3}\hat{p}
_{4}-y_{4}\hat{p}_{3} .\label{a}
\end{align}

The Poisson bracket can be written in terms of the $z_{i}$ variables as
\begin{align}
& \{f(z_{i},\bar{z}_{i}),g(z_{i},\bar{z}_{i})\}\notag \\
& = -2\mathrm{i} \omega \sum_{k=1}^{N} \sum_{j=2k-1}^{2k} n_k \bigg(\frac{\partial f}{\partial z_{j}}\frac{\partial g}{\partial \bar{z}_{j}}-
\frac{\partial f}{\partial \bar{z}_{j}}\frac{\partial g}{\partial z_{j}}\bigg)\label{Poisson}
\end{align}%
(in this Section we have $N=2$).

Functions of $z_k,\bar{z}_k$ Poisson commuting with $L_{12}$ and $L_{34}$ must satisfy
\begin{align}
z_{2}\partial _{z_{1}}f-z_{1}\partial _{z_{2}}f+\bar{z}_{2}\partial _{\bar{z}
_{1}}f-\bar{z}_{1}\partial _{\bar{z}_{2}}f & =0, \notag \\
z_{4}\partial _{z_{3}}f-z_{3}\partial _{z_{4}}f+\bar{z}_{4}\partial _{\bar{z}
_{3}}f-\bar{z}_{3}\partial _{\bar{z}_{4}}f &=0 \label{atri}.
\end{align}
A basis for the corresponding $SO(2)\times SO(2) $ invariants is given by
\begin{align}
\xi_1 & =z_{1}^{2}+z_{2}^{2},\quad \bar{\xi}_1=\bar{z}_{1}^{2}+\bar{z}_{2}^{2},\quad \eta_1= z_{1}\bar{z}_{1}+z_{2}\bar{z}_{2}, \notag \\
\xi_3 & =z_{3}^{2}+z_{4}^{2},\quad \bar{\xi}_3=\bar{z}_{3}^{2}+\bar{z}_{4}^{2},\quad \eta_2= z_{3}\bar{z}_{3}+z_{4}\bar{z}_{4},  \label{inv}
\end{align}%
Finally the integrals of motion must satisfy
\begin{equation}\label{posledni}
\{ H_4, f(\xi_1,\bar{\xi}_1,\eta_1,\xi_3,\bar{\xi}_3,\eta_2) \}=0.
\end{equation}

Solutions of equation (\ref{posledni}) are for instance
\begin{align}
E_1 &=\frac{1}{2} (|z_1|^2+|z_2|^2), \quad E_2 =\frac{1}{2} (|z_3|^2+|z_4|^2), \notag \\
Q_1 & = ({z}_1^2+z_2^2)^{n_2} (\bar{z}_3^2+\bar{z}_4^2)^{n_1},\label{b} \\
\bar{Q}_1 &= (\bar{z}_1^2+\bar{z}_2^2)^{n_2} (z_3^2+z_4^2)^{n_1}, \notag  \\
I_1 &= (z_1^2+z_2^2)(\bar{z}_1^2+\bar{z}_2^2), \,  I_2 = (z_3^2+z_4^2)(\bar{z}_3^2+\bar{z}_4^2).\notag
\end{align}
Only five of these integrals are functionally independent.

%These integrals are not all functionally independent. Indeed we have
%$$ I_1= H_1^2-4 n_1^2 \omega^2 L_{12}^2$$
%and similarly for $I_2$.

\subsection{Reduction of the first integrals}

The reduction is performed using the change of variables (\ref{polar}), (\ref{polarmomenta}) and the convention (\ref{17prime}). The integrals (\ref{a}) reduce to constants $L_{12}=\sqrt{k_1}, \, L_{34}=\sqrt{k_2}$. The integrals  (\ref{b}) reduce to nontrivial integrals for the Hamiltonian in Eq. (\ref{c}), namely
\begin{align}
E_1 &=\frac{1}{2} p_1^2+ \frac{k_1}{2x_1^2}+\frac{1}{2} n_1^2 \omega^2 x_1^2, \notag \\
E_2 &=\frac{1}{2} p_2^2+ \frac{k_2}{2x_2^2}+\frac{1}{2} n_2^2 \omega^2 x_2^2 \notag \\
\notag  Q_1 &= ( p_1^2+\frac{k_1}{x_1^2}-n_1^2 \omega^2 x_1^2 - 2 \mathrm{i} n_1 \omega p_1 x_1 )^{n_2}  \\
& \times ( p_2^2+\frac{k_2}{x_2^2}-n_2^2 \omega^2 x_2^2 + 2 \mathrm{i}  n_2 \omega p_2 x_2 )^{n_1},  \\
\notag  \bar{Q}_1 & = ( p_1^2+\frac{k_1}{x_1^2}-n_1^2 \omega^2 x_1^2 + 2 \mathrm{i}  n_1 \omega p_1 x_1 )^{n_2}  \\
&  \times ( p_2^2+\frac{k_2}{x_2^2}-n_2^2 \omega^2 x_2^2 - 2 \mathrm{i}  n_2 \omega p_2 x_2 )^{n_1}.\notag
\end{align}
The remaining two integrals in (\ref{b}) give nothing new and we have
\be
I_1 =4 (E_1^2-k_1 n_1^2 \omega^2), \quad  I_2 = 4 (E_2^2-k_2 n_2^2 \omega^2).
\ee

Three functionally independent real integrals of motion of the system with Hamiltonian (\ref{c})
can be chosen to be
\be
\{ E_1,E_2,Q=\frac{1}{2}(Q_1+\bar{Q}_1 )\}.
\ee
They are of order $2$, $2$ and $2(n_1+n_2) $ in the momenta, respectively. Their existence is the proof
of the maximal superintegrability of the considered system.

The integral of motion  $Q$ simplifies to give a second order one in two cases (that were known previously \cite{FMSUW}, \cite{Winternitz}).
They are
\begin{description}
\item{I)} $n_1=n_2=1$
\be\label{d}
\frac{4 E_1 E_2 - Q}{2 \omega^2} = (p_1 x_2-p_2 x_1)^2+\frac{k_1 x_2^2}{x_1^2}+\frac{k_2 x_1^2}{x_2^2}.
\ee
\item{II)} $n_1=1,n_2=2,k_2=0 $
\begin{align}\label{e}
&\left( \frac{8 E_1^2 E_2-Q}{8 \omega^2} -k_1E_2\right)^{1/2} \notag \\ &\qquad = p_1(x_2 p_1-x_1 p_2)-\omega^2 x_1^2 x_2+k_1 \frac{x_2}{x_1^2}
\end{align}
\end{description}
The integrals (\ref{d}) and (\ref{e}) are responsible for the separation of variables in polar and parabolic coordinates, respectively.
The integrals $\{ E_1, E_2 \} $ are responsible for the separation in cartesian coordinates.

\section{III. The general case}

Within the same approach, it is easy to extend the previous
picture to the general situation of a reduction from a $2N$ to a $N$%
--dimensional configuration space:

\begin{equation}
H_{2N}=\frac{1}{2}\sum_{i=1}^{2N}\hat{p}_{i}^{2}+\frac{\omega ^{2}}{2}%
\sum_{j=1}^{N}n_{j}^{2}(y_{2j-1}^{2}+y_{2j}^{2})\text{.}
\end{equation}%
Indeed, let us introduce the affine variables
\begin{equation*}
z_{k}=\hat{p}_{k}-{\mathrm{i}}\,n_{k}\omega y_{k},\quad k=1,\ldots ,2N
\end{equation*}%
so that the Hamiltonian reads%
\begin{equation*}
H_{2N}=\frac{1}{2}\sum_{k=1}^{2N}|z_{k}|^{2}\text{.}
\end{equation*}%
The Poisson bracket is defined as in eq. (\ref{Poisson}).
The invariants under the $SO(2)\times  \cdots \times SO(2)$ group action generated by $L_{12},\cdots
,L_{2N-1,2N}$ are
\begin{align}
\xi_{2k-1} & =z_{2k-1}^2+z_{2k}^2, \qquad k=1,2,\ldots,N, \notag \\
\bar{\xi}_{2k-1} & =\bar{z}_{2k-1}^2+\bar{z}_{2k}^2
\end{align}
apart from the quantities $L_{12},\cdots ,L_{2N-1,2N}$ and the \textquotedblleft 2-plane energies\textquotedblright\ which commute with
the Hamiltonian $H_{2N}$
\begin{equation}
|z_{1}|^{2}+|z_{2}|^{2},\ldots ,|z_{2N-1}|^{2}+|z_{2N}|^{2}.
\end{equation}%
Imposing
\begin{equation}
\{H_{2N},f(\xi,\,\bar{\xi})\}=0
\end{equation}%
where $\xi=(\xi_1,\ldots,\xi_{2N-1})$, $\bar{\xi}=(\bar{\xi_1},\ldots,\bar{\xi}_{2N-1})$, we get the differential equation
\begin{equation}
\sum_{k=1}^{N} n_k \bigg(\xi_{2k-1} \frac{\partial}{\partial \xi_{2k-1}} -\bar{\xi}_{2k-1} \frac{\partial}{\partial \bar{\xi}_{2k-1}}\bigg) f =0.
\end{equation}%
Its general solution depends on $2N-1$ invariants, which can be chosen as:
\begin{align}
Q_{2k-1}=&(z^2_{2k-1}+z^2_{2k})^{n_N} (\bar{z}^2_{2N-1}+\bar{z}^2_{2N})^{n_k},\notag\\ &\qquad k=1,\ldots, N-1,
\end{align}%
$\bar{Q}_{2k-1}$ and $I=|z^2_1+z^2_2|$.

Using the transformation (\ref{polar}) we now reduce the original
Hamiltonian to the following one:
\begin{equation}
H_{N}=\frac{1}{2}\sum_{i=1}^{N}p_{i}^{2}+\frac{1}{2}\sum_{i=1}^{N}\frac{k_{i}}{x_{i}^{2}%
}+\frac{\omega ^{2}}{2}\sum_{i=1}^{N}n_{i}^{2}x_{i}^{2}
\end{equation}%
The corresponding reduced invariants are:
\begin{align}
E_l &=\frac{1}{2} p_l^2+ \frac{k_l}{2x_l^2}+\frac{1}{2} n_l^2 \omega^2 x_l^2, \qquad l=1,\ldots,N, \notag \\
 R_{2l-1} &=\frac{1}{2} \left( Q_{l}+\bar{Q}_{l} \right) , \qquad l=1,\ldots,N-1
 \end{align}
 where
 \begin{align*}
Q_l &= ( p_l^2+\frac{k_l}{x_l^2}-n_l^2 \omega^2 x_l^2 - 2 i n_l \omega p_l x_l )^{n_N}  \\
& \times ( p_N^2+\frac{k_N}{x_N^2}-n_N^2 \omega^2 x_N^2 + 2 i n_N \omega p_N x_N )^{n_l}.
\end{align*}
There are $2N-1$ functionally independent integrals and consequently the system is maximally superintegrable,
proving the conjecture of \cite{Verrier}.

We have thus added a new maximally superintegrable system in $N$
dimensions to previously known ones (see, e.g., \cite{Evans}, \cite{Evans2}, \cite{Kalnins}, \cite{RW}, \cite{TWR}).

\section{IV. Open problems}

From the previous considerations, it emerges that it would be desirable to construct
systematically transformations mapping a superintegrable system into another
system, that is also superintegrable, and defined in a reduced phase space. It seems
natural to associate such transformations to the rich symmetry structure
possessed by superintegrable systems. For instance, changes of variables of
the type (\ref{polar}) are clearly related to invariance properties under
rotation. From this point of view, the role of higher order groups of
transformations generated by the flow associated to integrals that are polynomials in the momenta remains to be fully investigated. A quantum mechanical version of
this reduction procedure is also to be understood.
For $N=3$ the quantum system was treated in \cite{EvansII}. The reduction was performed for the classical system. The reduced system was then quantized in cartesian coordinates.

\section{Aknowledgements}

The authors wish to thank A. Ibort, G. Marmo, G. Rastelli, L. \v Snobl and G. Tondo for useful discussions. We also thank the anonymous referees for helpful suggestions. The research of P.W. was partly supported by NSERC of Canada. He also thanks the Facultad de F\'{\i}sicas, Universidad Complutense of Madrid, for hospitality and the Ministry of Education of Spain
for support during the realization of this project.
The research of M.A.R. was supported by the DGI under grant no. FIS2005-00752 and by the Universidad Complutense and the DGUI under grant no. GR74/07-910556.

\end{document}